\documentclass[aps,showpacs,nofootinbib,superscriptaddress]{revtex4}
\usepackage{epsf}
\usepackage{graphicx}
\usepackage{amsmath}
\newcommand{\ii}{\'\i}
\begin{document}

\title{ Can One Distinguish $\tau-$Neutrinos from Antineutrinos in
  Neutral-Current Pion Production Processes? } 

\author{ E. Hern\'andez} \affiliation{Grupo de F\'\i sica Nuclear,
  Departamento de F\ii sica Fundamental e IUFFyM,
Facultad de Ciencias, E-37008 Salamanca, Spain.}
\author{J.~Nieves} 
\affiliation{Departamento de F\'\i sica At\'omica, Molecular y Nuclear,
\\ Universidad de Granada, E-18071 Granada, Spain}
\author{M.~Valverde} 
\affiliation{Departamento de F\'\i sica At\'omica, Molecular y Nuclear,
\\ Universidad de Granada, E-18071 Granada, Spain}

\pacs{12.15.Mm,25.30.Pt,13.15.+g}

\begin{abstract}
 A potential way to distinguish $\tau-$neutrinos from antineutrinos,
 below the $\tau-$production threshold, but above the pion production
 one, is presented. It is based on the different behavior of the
 neutral current pion production off the nucleon, depending on whether
 it is induced by neutrinos or antineutrinos. This procedure for
 distinguishing $\tau-$neutrinos from antineutrinos neither relies on
 any nuclear model, nor it is affected by any nuclear effect
 (distortion of the outgoing nucleon waves, etc...).  We show that
 neutrino-antineutrino asymmetries occur both in the totally
 integrated cross sections and in the pion azimuthal differential
 distributions. To define the asymmetries for the latter distributions
 we just rely on Lorentz-invariance.  All these asymmetries are
 independent of the lepton family and can be experimentally measured
 by using electron or muon neutrinos, due to the lepton family
 universality of the neutral current neutrino interaction. Nevertheless and to
 estimate their size, we have also used the chiral model of {\it
 hep-ph/0701149}  at intermediate energies. Results are really
 significant since the differences between neutrino and antineutrino
 induced reactions are always large in all physical channels.

\end{abstract}

\maketitle

\section{Introduction}

Distinguishing $\tau-$neutrinos from antineutrinos, can be easily done
above the $\tau-$production threshold energy, since then the
charged--current (CC) channel is open. Conservation of lepton number
implies that neutrinos generate a tau lepton, while an antineutrino
produces an antilepton. The charge of the tau lepton unambiguously
reveals the nature of the incident neutrino beam. Below the
$\tau-$threshold, $\tau-$neutrinos or antineutrinos only interact with
matter via neutral--current (NC) driven processes, and the outgoing lepton
is also a neutrino or antineutrino, and the difference does not
manifest itself in this clear way. 

Nonetheless, it is out of any doubt the great interest of
distinguishing between neutrinos and antineutrinos in different
scenarious: CP--violation in $\nu_\mu$--$\nu_\tau$
oscillations~\cite{HHW02}, supernova--neutrino 
physics~\cite{BRJK03}\footnote{ At
typical supernova energies, $\nu_\tau$ neutrinos do not participate in
CC reactions}, \ldots

Very recently, Jachowicz and collaborators suggested that in
neutrino--nucleus processes, the helicity--related differences between
neutrino and antineutrinos induce some asymmetries in the
polarization of the ejected nucleons~\cite{JVRH04}. From this fact,
the authors of this reference conclude that these asymmetries
represent a potential way to distinguish  neutrinos from
antineutrinos in NC neutrino scattering on nuclei.

 We present in this letter an alternative manner to distinguish
 $\tau-$neutrinos from antineutrinos, below the $\tau-$production
 threshold ($\approx$ 3.5 GeV in the Laboratory (LAB) frame for
 production off the nucleon), but
 above the pion production one ($\approx$ 0.15 GeV in the LAB
 frame). The method is based on the different behavior of the NC pion
 production reaction in the nucleon, depending on whether it is
 induced by neutrinos or antineutrinos. This procedure for
 distinguishing $\tau-$neutrinos from antineutrinos neither relies on
 any nuclear model, nor it is affected by any nuclear effect
 (distortion of the outgoing nucleon waves, etc...).  We show that
 neutrino-antineutrino asymmetries occur both in the totally
 integrated cross sections and in the pion azimuthal differential
 distributions. To define the asymmetries for the latter distributions
 we just rely on Lorentz-invariance, and experimentally it requires
 determining the neutrino scattering plane, detecting either a neutral
 or a positively charged pion and measuring its momentum. Since the
 outgoing neutrino will  not likely be  detected, to fix the neutrino
 scattering plane would also require measuring the outgoing nucleon
 momentum and some knowledge of the incoming neutrino momentum.

 All these asymmetries are independent of the lepton family and can be
experimentally measured by using electron or muon neutrinos, due to
the lepton family universality of the NC neutrino
interaction. Nevertheless and to estimate their size, we have used the
chiral model of Ref.~\cite{HNV07} to compute them 
 up to neutrino/antineutrino energies of 2 GeV.

\section{Kinematics and cross sections}

\begin{figure}[tbh]
\centerline{\includegraphics[height=5cm]{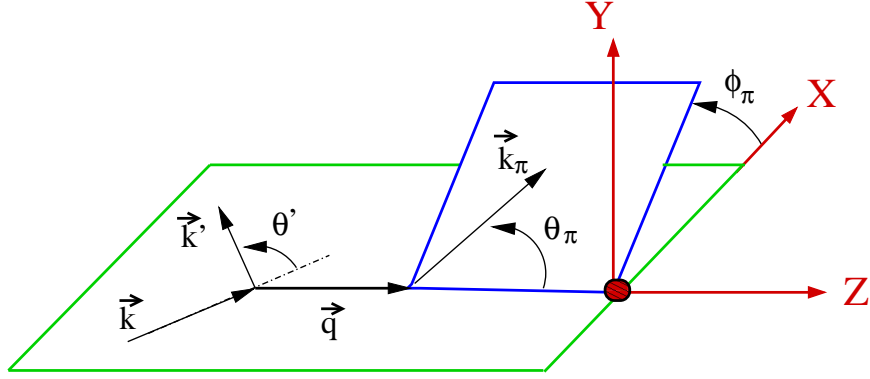}}
\caption{\footnotesize Definition of the different kinematical
variables used through this work.}\label{fig:coor}
\end{figure}

We will focus on the neutrino--pion production reaction off the nucleon
driven by neutral currents,
\begin{equation}
  \nu_l (k) +\, N(p)  \to \nu_l (k^\prime) + N(p^\prime) +\, \pi(k_\pi) 
\label{eq:reac}
\end{equation}
though the generalization of the obtained expressions to antineutrino
induced reactions is straightforward.

The unpolarized differential cross section, with respect to the
outgoing neutrino and pion kinematical variables is given in the
LAB frame (kinematics is sketched in Fig~\ref{fig:coor})
by\footnote{To obtain Eq.~(\ref{eq:sec}) we have neglected the
four-momentum carried out by the intermediate $Z-$boson with respect
to its mass $(M_Z)$, and have used the existing relation between the
gauge weak coupling constant, $g = e/\sin \theta_W$, and the Fermi
constant: $G/\sqrt 2 = g^2/8M^2_W$, with $e$ the electron charge,
$\theta_W$ the Weinberg angle, $\cos\theta_W=M_W/M_Z$ and $M_W$ the
$W-$boson mass.}

\begin{equation}
\frac{d^{\,5}\sigma_{\nu}}{d\Omega(\hat{k^\prime})dE^\prime 
d\Omega(\hat{k}_\pi) } =
\frac{|\vec{k}^\prime|}{|\vec{k}~|}\frac{G^2}{16\pi^2} 
 \int_0^{+\infty}\frac{dk_\pi k_\pi^2}{E_\pi}
 L_{\mu\sigma}^{(\nu)}\left(W^{\mu\sigma}_{{\rm NC} \pi}\right)^{(\nu)} 
 \label{eq:sec}
\end{equation}
with $\vec{k}$ and $\vec{k}^\prime~$ the LAB neutrino momenta,
$E^{\prime}= |\vec{k}^{\prime\,}|$, the energy of the outgoing
neutrino, $\vec{k}_\pi$ and $E_\pi$, the momentum and energy of the
pion in the LAB system, $G=1.1664\times 10^{-11}$ MeV$^{-2}$, the
Fermi constant and $L$ and $W$ the leptonic and hadronic tensors,
respectively. The leptonic tensor is given by (in our convention, we
take $\epsilon_{0123}= +1$ and the metric $g^{\mu\nu}=(+,-,-,-)$):
\begin{eqnarray}
L_{\mu\sigma}^{(\nu)}&=& (L^{(\nu)}_s)_{\mu\sigma}+ {\rm i}
 (L^{(\nu)}_a)_{\mu\sigma} =
 k^\prime_\mu k_\sigma +k^\prime_\sigma k_\mu
- g_{\mu\sigma} k\cdot k^\prime + {\rm i}
\epsilon_{\mu\sigma\alpha\beta}k^{\prime\alpha}k^\beta \label{eq:lep}
\end{eqnarray}
and it is orthogonal to $q^\mu=(k-k^\prime)^\mu$ for massless neutrinos, i.e,
$L_{\mu\sigma}^{(\nu)} q^\mu = L_{\mu\sigma}^{(\nu)} q^\sigma = 0$.

The hadronic tensor includes all sort of non-leptonic
vertices and it reads
\begin{eqnarray}
(W^{\mu\sigma}_{{\rm NC} \pi})^{(\nu)} &=& \frac{1}{4M}\overline{\sum_{\rm
 spins}} \int\frac{d^3p^\prime}{(2\pi)^3} \frac{1}{2E^\prime_N}
  \delta^4(p^\prime+k_\pi-q-p) \langle N^\prime \pi |
 j^\mu_{\rm nc}(0) | N \rangle \langle N^\prime \pi | j^\sigma_{\rm nc}(0) | N
 \rangle^*
\label{eq:wmunu}
\end{eqnarray}
with $M$ the nucleon mass  and $E^\prime_N$  the energy
of the outgoing nucleon.  In the  sum over
initial and final nucleon spins\footnote{Since right-handed neutrinos
are sterile, only left-handed neutrinos are considered.}, the bar over
the sum denotes the average over the initial ones.  As for  the one particle
states they are normalized so that $\langle \vec{p}\, | \vec{p}^{\,\prime}
\rangle = (2\pi)^3 2p_0 \delta^3(\vec{p}-\vec{p}^{\,\prime})$, and
finally for the neutral current we take
\begin{equation}
j^\mu_{\rm nc} = \overline{\Psi}_u\gamma^\mu(1-\frac83 \sin^2\theta_W-
\gamma_5)\Psi_u  - \overline{\Psi}_d\gamma^\mu(1-\frac43 \sin^2\theta_W-
\gamma_5)\Psi_d  - \overline{\Psi}_s\gamma^\mu(1-\frac43 \sin^2\theta_W-
\gamma_5)\Psi_s  
\end{equation}
with $\Psi_u$, $\Psi_d$ and $\Psi_s$ quark fields, and $\theta_W$ the
Weinberg angle ($\sin^2\theta_W= 0.231$). Note that with all these
definitions, the matrix element $\langle N^\prime \pi | j^\mu_{\rm
nc}(0) | N \rangle$ is dimensionless. Both, lepton and hadron tensors
are independent of the neutrino lepton family, and therefore the cross
section for the reaction of Eq.~(\ref{eq:reac}) is the same for
electron, muon or tau incident neutrinos.  As the quantity 
$L_{\mu\sigma}^{(\nu)}\left(W^{\mu\sigma}_{{\rm
NC} \pi}\right)^{(\nu)}$ is a Lorentz scalar, to evaluate it we take for
convenience $\vec{q}$ in the $Z$ direction (see Fig.~\ref{fig:coor}).
Referring now the pion variables to the outgoing $\pi N$ pair center
of mass frame (as it is usual in pion electroproduction) would be
readily done by means of a boost in the $Z$ direction. Note that the
azimuthal angle $\phi_\pi$ is left unchanged by such a boost.

For antineutrino induced reactions we have 
\begin{equation}
L_{\mu\sigma}^{(\bar\nu)} = L_{\sigma\mu}^{(\nu)}, \qquad
(W^{\mu\sigma}_{{\rm NC} \pi})^{(\bar\nu)} = 
(W^{\mu\sigma}_{{\rm NC} \pi})^{(\nu)} \label{eq:anti}
\end{equation}
For the sake of simplicity, from now on, we will omit in the hadronic
tensor the explicit reference to $\nu$ or $\bar\nu$ and the NC $\pi$
label. By construction, the hadronic tensor accomplishes
\begin{eqnarray}
W^{\mu\sigma}= W^{\mu\sigma}_s + {\rm i} W^{\mu\sigma}_a 
\end{eqnarray}
with $W^{\mu\sigma}_s$ ($W^{\mu\sigma}_a$) real symmetric
(antisymmetric) tensors.  The hadronic tensor is completely
determined by up to a total of 19  Lorentz
scalar and real, structure functions $W_i(q^2,\, p\cdot q,\, p\cdot
k_\pi,\, k_\pi \cdot q)$,
\begin{eqnarray}
  W^{\mu\nu}_{s,a} &=& \left(W^{\mu\nu}_{s,a}\right)^{\rm PC} +
  \left(W^{\mu\nu}_{s,a}\right)^{\rm PV} \label{eq:w1}\\ &&\nonumber\\
  \left(W^{\mu\nu}_s\right)^{\rm PC} &=& W_1 g^{\mu\nu} + W_2 p^\mu
  p^\nu + W_3 q^\mu q^\nu + W_4 k^\mu_\pi k^\nu_\pi + W_5 (q^\mu p^\nu
  + q^\nu p^\mu )+ W_6 (q^\mu k_\pi^\nu + q^\nu k_\pi^\mu )\nonumber
  \\ &+&W_7(p^\mu k_\pi^\nu + p^\nu k_\pi^\mu )  \\
  \left(W^{\mu\nu}_s\right)^{\rm PV} &=& W_8
  \left(q^\mu\epsilon^\nu_{.\alpha\beta\gamma}k_\pi^\alpha p^\beta
  q^\gamma + q^\nu\epsilon^\mu_{.\alpha\beta\gamma}k_\pi^\alpha p^\beta
  q^\gamma \right) + W_9
  \left(p^\mu\epsilon^\nu_{.\alpha\beta\gamma}k_\pi^\alpha p^\beta
  q^\gamma + p^\nu\epsilon^\mu_{.\alpha\beta\gamma}k_\pi^\alpha p^\beta
  q^\gamma \right)\nonumber\\ &+&
  W_{10}\left(k_\pi^\mu\epsilon^\nu_{.\alpha\beta\gamma}k_\pi^\alpha
  p^\beta q^\gamma +
  k_\pi^\nu\epsilon^\mu_{.\alpha\beta\gamma}k_\pi^\alpha p^\beta
  q^\gamma \right)\\ &&\nonumber \\ 
\left(W^{\mu\nu}_a\right)^{\rm PV}
  &=& W_{11} (q^\mu p^\nu - q^\nu p^\mu )+ W_{12} (q^\mu k_\pi^\nu -
  q^\nu k_\pi^\mu ) +W_{13}(p^\mu k_\pi^\nu - p^\nu k_\pi^\mu ) \\
  \left(W^{\mu\nu}_a\right)^{\rm PC} &=& W_{14}
  \epsilon^{\mu\nu\alpha\beta}p_\alpha q_\beta + W_{15}
  \epsilon^{\mu\nu\alpha\beta}p_\alpha k_{\pi\beta} +W_{16}
  \epsilon^{\mu\nu\alpha\beta}q_\alpha k_{\pi\beta}+W_{17}
  \left(q^\mu\epsilon^\nu_{.\alpha\beta\gamma}k_\pi^\alpha p^\beta
  q^\gamma - q^\nu\epsilon^\mu_{.\alpha\beta\gamma}k_\pi^\alpha p^\beta
  q^\gamma \right)\nonumber\\ &+&
  W_{18}\left(p^\mu\epsilon^\nu_{.\alpha\beta\gamma}k_\pi^\alpha
  p^\beta q^\gamma - p^\nu\epsilon^\mu_{.\alpha\beta\gamma}k_\pi^\alpha
  p^\beta q^\gamma \right)+
  W_{19}\left(k_\pi^\mu\epsilon^\nu_{.\alpha\beta\gamma}k_\pi^\alpha
  p^\beta q^\gamma -
  k_\pi^\nu\epsilon^\mu_{.\alpha\beta\gamma}k_\pi^\alpha p^\beta
  q^\gamma \right) \label{eq:w2}
\end{eqnarray}
Though $W^{\mu\nu}$ is not orthogonal to $q^\mu$, the
$W_3$, $W_5$, $W_6$, $W_8$, $W_{11}$, $W_{12}$ and $W_{17}$ terms do
not contribute to the differential cross section since the leptonic
tensor is  orthogonal to the four vector $q^\mu$ for massless
neutrinos.

The tensor $\left(W^{\mu\nu}\right)^{\rm
PV}=\left(W^{\mu\nu}_s\right)^{\rm PV} +~ {\rm i}
\left(W^{\mu\nu}_a\right)^{\rm PV}$  when contracted with the leptonic
one, $L_{\mu\nu}^{(\nu)}$, provides a pseudo-scalar quantity,
i.e., such contraction is not invariant under a parity
transformation. Indeed, under a parity transformation we have, 
\begin{equation}
L_{\mu\nu}^{(\nu)} \to \left(L^{\nu\mu}\right)^{(\nu)}, \qquad \left(W_{\mu\nu}\right)^{\rm
PV} \to -\left(W^{\nu\mu}\right)^{\rm PV} \label{eq:parity}
\end{equation}
whereas the tensor $\left(W^{\mu\nu}\right)^{\rm
PC}=\left(W^{\mu\nu}_s\right)^{\rm PC} +~ {\rm i}
\left(W^{\mu\nu}_a\right)^{\rm PC}$ transforms as $\left
(L^{\mu\nu}\right)^{(\nu)}$. This explains the origin of the adopted
labels PC and PV, which stand for parity violating and conserving
contributions to the fifth differential cross section
$d^{\,5}\sigma/d\Omega(\hat{k^\prime})dE^\prime d\Omega(\hat{k}_\pi)$,
respectively\footnote{Similar conclusions can be also reached by just 
noting that the symmetric part of the leptonic tensor does not contain
any Levi-Civita $\epsilon^{\mu\nu\alpha\beta}$ pseudo--tensor, whereas
its antisymmetric part is just given by a Levi-Civita tensor. Thus, to get a
parity invariant quantity, the symmetric part of the hadronic tensor
should not have any Levi-Civita tensor, while
all terms of its antisymmetric part should  contain a
Levi-Civita tensor.}. The triple differential cross section
$d^{\,3}\sigma/d\Omega(\hat{k^\prime})dE^\prime$ is a scalar, up to
the factor $ |\vec{k}^\prime|/|\vec{k}~|$. Thus all
parity-violating contributions must disappear when performing the 
pion solid angle integration. Note that the coordinate system used to
define $d\Omega(\hat{k}_\pi)$ involves the pseudo-vector $\vec{k}
\times \vec{k}^\prime$ to set up the $Y-$axis, which induces the
non-parity invariant nature of
$d^{\,5}\sigma/d\Omega(\hat{k^\prime})dE^\prime d\Omega(\hat{k}_\pi)$. In
electropion production processes, the leptonic tensor is purely
symmetric, and the symmetric part of the hadronic one can not contain
terms involving the  Levi-Civita tensor, since the
electromagnetic interaction preserves parity. Hence in that case
$d^{\,5}\sigma/d\Omega(\hat{k^\prime})dE^\prime d\Omega(\hat{k}_\pi)$
turns out to be a scalar under parity.

A final remark concerns the  time-reversal ($T$) violation
effects apparently encoded in the decomposition of the hadronic tensor in
Eqs.~(\ref{eq:w1}--\ref{eq:w2}). Under a time reversal transformation,
and taking into account the antiunitary character of the
$T-$operator,  we have
\begin{equation}
L_{\mu\nu}^{(\nu)} \to \left(L^{\mu\nu}\right)^{(\nu)}, \qquad
\left(W_{\mu\nu}\right)_{\rm PC} \to \left(W^{\mu\nu}\right)^{\rm PC},
\qquad \left(W_{\mu\nu}\right)^{\rm PV} \to
-\left(W^{\mu\nu}\right)^{\rm PV} \label{eq:time}
\end{equation}
and therefore $L^{(\nu)}_{\mu\nu} W^{\mu\nu}$ is not $T-$invariant
either, because of the presence of the PV terms in the hadronic
tensor\footnote{Note that transformations given in
Eqs.~(\ref{eq:anti}), (\ref{eq:parity}) and (\ref{eq:time}) imply
that the leptonic tensor, by itself, is invariant under CPT.}. This
does not necessarily means that there exists a violation of
$T-$invariance in the process~\cite{KLS68}. Time reversal is the transformation
that changes the direction of time: $t\to -t$. Thus invariance under
time reversal means that the description of physical processes does
not show any asymmetry if we look backward in time. The invariance
under time reversal is equivalent to 
\begin{equation}
|M_{i\to f}|^2 = |M_{Tf\to Ti}|^2 
\end{equation}
where $i \to f$ denotes the transition from the initial state $i$ to
the final state $f$, being $M_{i\to f}$ its corresponding transition
amplitude, and $Tf$ and $Ti$ denote the states obtained from $f$ and
$i$, respectively, by reversing the momenta, spins, etc.. When one is
dealing with electromagnetic or weak interactions, which may be
treated to first order in the interaction Hamiltonian ($H_I$), the
transition matrix operator can be approximated by $H_I$, being it then
hermitian. In these circumstances $|M_{Tf\to Ti}|\approx |M_{Ti\to
Tf}|$ and the invariance under time reversal may be written in the
following form
\begin{equation}
|M_{i\to f}|^2 = |M_{Tf\to Ti}|^2 \approx |M_{Ti\to Tf}|^2 \label{eq:time_approx}
\end{equation}
The above equation implies that the transition $i\to f$ cannot show up
correlations that change sign under time reversal ($T-$odd
correlations). Thus at first order in the interaction Hamiltonian, the
tensor $\left(W^{\mu\nu}\right)^{\rm PV}$ should vanish, since it
would lead to violations of time reversal invariance. However, here we
have strong interacting final states (pion and nucleon) and in fact
the transition matrix operator is not hermitian, since it cannot be
approximated by $H_I$. Therefore Eq.~(\ref{eq:time_approx}) does not
hold and time reversal invariance does not forbid a non-vanishing
$\left(W^{\mu\nu}\right)^{\rm PV}$ tensor.  In summary, besides
genuine time reversal violations~\cite{CLS70}\footnote{One is faced
with the problem of eliminating final-state-interaction effects, or at
least of having them under control, in order to have a significant
test of $T$ violation~\cite{KLS68,CLS70}.}, the existence of strong
final state interaction effects\footnote{For instance, at intermediate energies
and for CC driven processes, the $\Delta(1232)$ resonance plays a
central role~\cite{Adler}. The inclusion of the resonance
width accounts partially for the strong final state interaction
effects.} allows for the existence
of $T-$odd correlations in $(L^{(\nu)})_{\mu\nu} W^{\mu\nu}$, induced
by the $\left(W^{\mu\nu}\right)^{\rm PV}$ term of the hadronic
tensor.

After this discussion, we are in conditions of studying the pion
azimuthal angle dependence of the differential cross section. With our
election of kinematics ($\vec{k},\vec{k}^\prime$ in the XZ plane), we
find $(L^{(\nu)}_s)_{0y} = (L^{(\nu)}_s)_{xy} =
(L^{(\nu)}_s)_{zy}=(L^{(\nu)}_a)_{0x}=(L^{(\nu)}_a)_{0z}=(L^{(\nu)}_a)_{xz}=0$,
and then
\begin{eqnarray}
\int_0^{+\infty}\frac{dk_\pi k_\pi^2}{E_\pi} (L^{(\nu)}_s)_{\mu\nu} 
W^{\mu\nu}_s &=& \int_0^{+\infty}\frac{dk_\pi k_\pi^2}{E_\pi} \Big\{ (L^{(\nu)}_s)_{00}W^{00}_s
+ 2(L^{(\nu)}_s)_{0x}W^{0x}_s + 
2(L^{(\nu)}_s)_{0z}W^{0z}_s + (L^{(\nu)}_s)_{xx}W^{xx}_s\nonumber \\ 
&+& (L^{(\nu)}_s)_{yy}W^{yy}_s + (L^{(\nu)}_s)_{zz}W^{zz}_s 
+  2 (L^{(\nu)}_s)_{xz}W^{xz}_s \Big \}\nonumber \\ 
 &=& A_s + B_s \cos\phi_\pi + C_s
\cos 2\phi_\pi+ D_s \sin\phi_\pi + E_s
\sin 2\phi_\pi \\
\int_0^{+\infty}\frac{dk_\pi k_\pi^2}{E_\pi}\ (L^{(\nu)}_a)_{\mu\nu} W^{\mu\nu}_a &=& 2 \int_0^{+\infty}\frac{dk_\pi k_\pi^2}{E_\pi}\Big\{ (L^{(\nu)}_a)_{0y}W^{0y}_a
+ (L^{(\nu)}_a)_{xy}W^{xy}_a + 
(L^{(\nu)}_a)_{yz}W^{yz}_a \Big\} \nonumber \\
&=& -A_a - B_a \cos\phi_\pi- D_a \sin\phi_\pi
\end{eqnarray}
where we explicitly show the $\phi_\pi$ dependence\footnote{All
structure functions, $W_{i=1,\cdots 19}$, depend on the Lorentz scalar
$p\cdot k_\pi$ and $k_\pi \cdot q$ factors, which are functions of the
angle formed between the $\vec{q}$ and $\vec{k}_\pi$ vectors, and thus
they are independent of $\phi_\pi$, when $\vec{q}$ is taken along the
$Z-$axis. }. The PV term of the hadronic tensor has led to the parity
violating $\sin\phi_\pi$ and $\sin 2\phi_\pi$ contributions (all of
them proportional to $k_{\pi y})$. They disappear when the pion solid
angle integration is performed, as anticipated. Thanks to
Eq.~(\ref{eq:anti}), we obtain for neutrino and antineutrino reactions
\begin{eqnarray}
(\nu, \nu)\to \frac{d^{\,5}\sigma_{\nu}}
{d\Omega(\hat{k^\prime})dE^\prime d\Omega(\hat{k}_\pi) } &=&
\frac{|\vec{k}^\prime|}{|\vec{k}~|}\frac{G^2}{16\pi^2} \Big\{ A_s+A_a +
    (B_s+B_a) \cos\phi_\pi + C_s \cos 2\phi_\pi +(D_s+D_a) \sin\phi_\pi 
\nonumber \\
&+& E_s \sin 2\phi_\pi 
    \Big\} \label{eq:nuphi}\\
(\bar\nu, \bar\nu)\to \frac{d^{\,5}\sigma_{\bar\nu}}
{d\Omega(\hat{k^\prime})dE^\prime d\Omega(\hat{k}_\pi) } &=&
\frac{|\vec{k}^\prime|}{|\vec{k}~|}\frac{G^2}{16\pi^2} \Big \{ A_s-A_a +
    (B_s-B_a) \cos\phi_\pi + C_s \cos 2\phi_\pi  +(D_s-D_a) \sin\phi_\pi 
\nonumber \\
&+& E_s \sin 2\phi_\pi 
    \Big\} \label{eq:nubarphi}
\end{eqnarray}

\section{Neutrino-antineutrino asymmetries}

From Eqs.~(\ref{eq:nuphi}) and~(\ref{eq:nubarphi}), we see
that neutrino-antineutrino asymmetries may occur both in the totally
integrated cross sections and in the pion azimuthal differential
distributions. For the neutrino and antineutrino induced
total cross sections we have
\begin{eqnarray}
(\nu, \nu)\to \sigma_{\nu} &=& \frac{G^2}{16\pi^2 |\vec{k}~|} \int
    d\Omega(\hat{k^\prime})dE^\prime d\Omega(\hat{k}_\pi)
    |\vec{k}^\prime|\Big (
    A_s+A_a \Big) \label{eq:a-nuphi}\\ 
    (\bar\nu, \bar\nu)\to
    \sigma_{\bar\nu} &=& \frac{G^2}{16\pi^2 |\vec{k}~|} \int
    d\Omega(\hat{k^\prime})dE^\prime d\Omega(\hat{k}_\pi)
    |\vec{k}^\prime| \Big ( A_s-A_a \Big ),
    \label{eq:a-nubarphi}
\end{eqnarray}
while for the azimuthal distributions, relations of the type 
\begin{eqnarray}
\frac{d^{\,5}\sigma}{d\Omega(\hat{k^\prime})dE^\prime
d\Omega(\hat{k}_\pi)}
(\phi_\pi)-\frac{d^{\,5}\sigma}{d\Omega(\hat{k^\prime})dE^\prime
d\Omega(\hat{k}_\pi)}
(\phi_\pi+\pi)\Big|_{\nu} &=&\left\{ (B_s+B_a)
\cos\phi_\pi+ (D_s+D_a)
\sin\phi_\pi \right\} \frac{|\vec{k}^\prime|}{|\vec{k}~|}\frac{G^2}{8\pi^2}
\label{eq:phisy} \\
\frac{d^{\,5}\sigma}{d\Omega(\hat{k^\prime})dE^\prime
d\Omega(\hat{k}_\pi)}
(\phi_\pi)-\frac{d^{\,5}\sigma}{d\Omega(\hat{k^\prime})dE^\prime
d\Omega(\hat{k}_\pi)}
(\phi_\pi+\pi)\Big|_{\bar\nu} &=& 
\left\{ (B_s-B_a) \cos\phi_\pi +  (D_s-D_a)
\sin\phi_\pi \right\}  \frac{|\vec{k}^\prime|}{|\vec{k}~|}\frac{G^2}{8\pi^2}
\label{eq:phiasy}
\end{eqnarray}
enhance neutrino-antineutrino asymmetries.  To increase the statistics in
these azimuthal asymmetries, it is interesting to study the
relations of Eqs.~(\ref{eq:phisy}) and~(\ref{eq:phiasy}) for
integrated cross sections. From the discussion in the
Appendix~\ref{app:piad}  it
follows
\begin{eqnarray}
\frac12\left(\frac{d\sigma(\phi_\pi)}{d\phi_\pi}-\frac{d\sigma(\phi_\pi+\pi)}{d\phi_\pi}\right)\Big|_{\nu} &=& 
({\cal B}_s+{\cal B}_a)
\cos\phi_\pi+ ({\cal D}_s+{\cal D}_a)
\sin\phi_\pi \label{eq:int-phisy}
 \\
\frac12\left(\frac{d\sigma(\phi_\pi)}{d\phi_\pi}-\frac{d\sigma(\phi_\pi+\pi)}{d\phi_\pi}\right)\Big|_{\bar\nu} &=& 
 ({\cal B}_s-{\cal B}_a) \cos\phi_\pi +  ({\cal D}_s-{\cal D}_a)
\sin\phi_\pi \label{eq:int-phiasy}
\end{eqnarray}
where for each event, $\phi_\pi$ is the angle formed by the neutrino
and pion scattering planes. Since the outgoing
neutrino will not likely be   detected, if the incoming neutrino
momentum is known, the neutrino scattering plane will be set up by
detecting also the outgoing nucleon. Indeed, the neutrino scattering
plane is also determined by the incoming neutrino and the transferred
momenta, and in the LAB system this latter one is given by the sum of
the outgoing nucleon and pion momenta ($\vec{q}=\vec{p}^{\,\prime} +
\vec{k}_\pi $). Thus the strategy will be detecting in coincidences
the outgoing pion and nucleon momenta, use those to determine the
neutrino scattering plane and then measure the corresponding
$\phi_\pi$ angle.

All these relations can be used to distinguish $\tau-$neutrinos from
antineutrinos, below the $\tau-$production threshold, but above the
pion production one.  Thus and using a proton target, the total cross section
or the dependence on the angle formed by the neutrino and pion planes
in either of the channels,
\begin{eqnarray}
\nu p \to \nu n \pi^+,  \quad \nu p \to \nu p \pi^0 \nonumber\\
\bar\nu p \to \bar\nu n \pi^+,  \quad \bar\nu p \to \bar\nu p \pi^0,
\end{eqnarray}
may be used to determine the nature of the incident $\tau-$neutrino beam below
the $\tau-$production threshold. 

Moreover, we would like to point out that the
$(A_s\pm A_a) $, $(B_s\pm B_a) $ and $(D_s\pm D_a)$ response
functions, the integrated cross sections $\sigma_{\nu}$,
$\sigma_{\bar\nu}$ and the partially integrated quantities $({\cal
B}_s\pm {\cal B}_a) $ and $({\cal D}_s\pm {\cal D}_a)$ are independent
of the lepton family and can be experimentally measured by using
electron or muon neutrinos, due to the lepton family universality of
the NC neutrino interaction.  Thus one would not have to rely on any
theoretical model to determine the asymmetry relations proposed in
this section.

\subsection{Neutrino-antineutrino asymmetries from the chiral model of
  Ref.~\cite{HNV07}}

The above suggestion for distinguishing neutrino from antineutrino in
 NC pion production reactions will be of no value if the terms $A_a,\,
 B_a$ and $D_a$ are much smaller than $A_s,\, B_s$ and $D_s$. To
 estimate these response functions and address this issue, we have used
 the model developed in Ref.~\cite{HNV07} to study the weak pion
 production off the nucleon at intermediate energies. In this model,
 besides the Delta pole mechanism $\Delta P$ (weak excitation of the
 $\Delta(1232)$ resonance and its subsequent decay into $N\pi$), some
 background terms, required by chiral symmetry, are also
 considered. These chiral background terms are calculated within a SU(2)
 non-linear $\sigma$ model involving pions and nucleons, which
 implements the pattern of spontaneous chiral symmetry breaking of
 QCD. The model gives a fair description of the neutrino and
 antineutrino CC and NC available data. Details can be found
 in Ref.~\cite{HNV07}.

\begin{figure}[tbh]
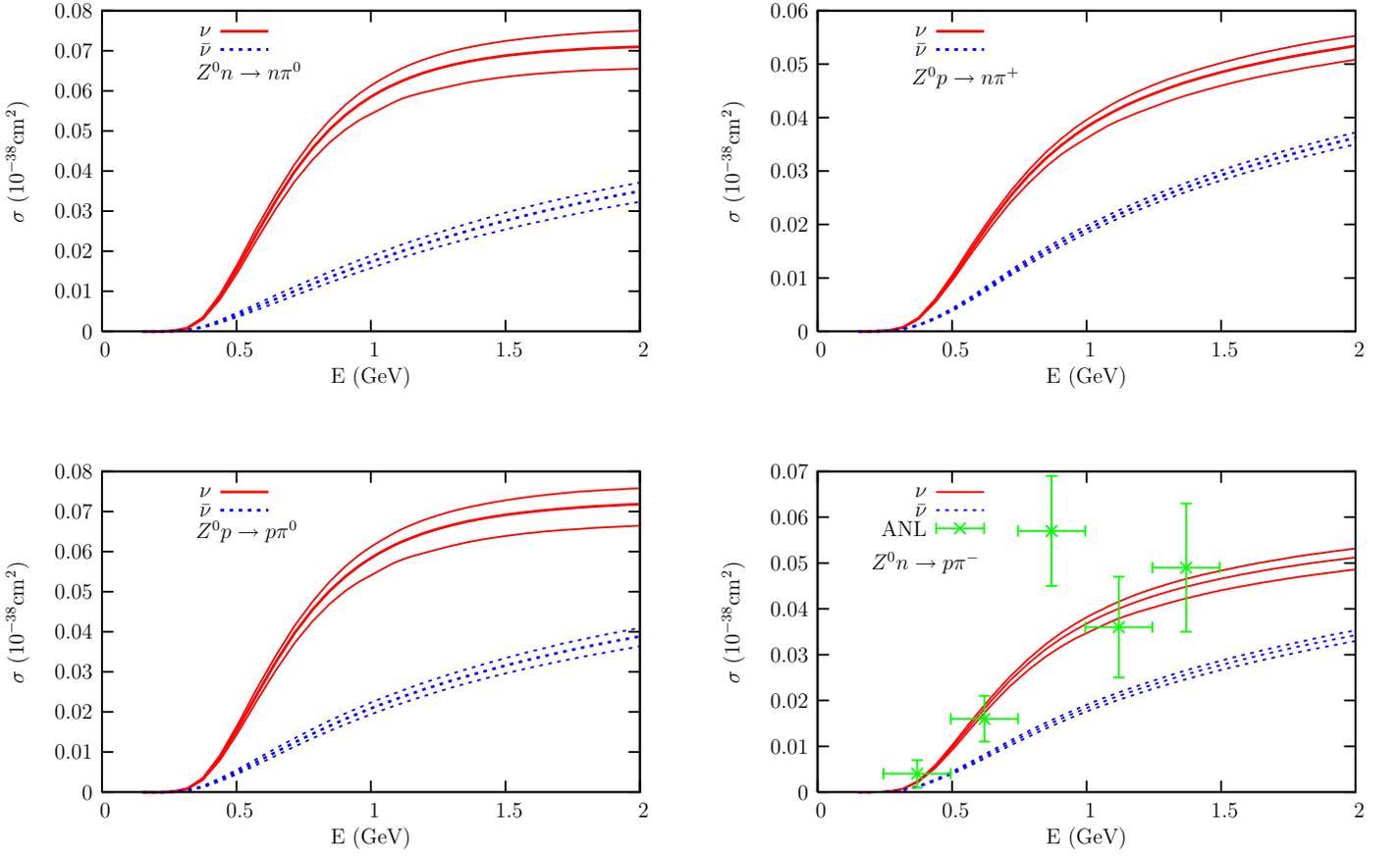

\begin{center}
\makebox[0pt]{\includegraphics[scale=0.7]{npi0.epsi}\hspace{1cm}\includegraphics[scale=0.7]{npiM.epsi}}\\\vspace{1cm}
\makebox[0pt]{\includegraphics[scale=0.7]{ppi0.epsi}\hspace{1cm}\includegraphics[scale=0.7]{ppim.epsi}}
\end{center}
\caption{\footnotesize Neutrino and antineutrino NC pion production
  total cross sections predicted by the model of Ref.~\cite{HNV07}, as
  a function of the incoming neutrino or antineutrino energy. An invariant
  pion-nucleon mass cut of $W \le 1.4$ GeV has been
  implemented. Central lines stand for the result of the model.  68\%
  CL bands (external lines), inferred from uncertainties of the model,
  are also displayed.  Experimental cross sections for neutrino $\pi^-$
  production are taken from Ref.~\cite{anl1NC}.}\label{fig:sectot}
\end{figure}

 In Fig.~\ref{fig:sectot}, we show neutrino and antineutrino NC pion
 production total cross sections predicted by the model of
 Ref.~\cite{HNV07}, as a function of the incoming neutrino or
 antineutrino energy.
 There, 68\% CL bands (external lines), inferred from uncertainties of
 the model of Ref.~\cite{HNV07} are also displayed. Those
 uncertainties come from re-adjusting the $C_5^A(q^2)$ form--factor
 that controls the largest term of the $\Delta-$axial contribution. On
 the other hand, since the main dynamical ingredients of the model are
 the excitation of the  $\Delta$ resonance and the non-resonant
 contributions deduced from the leading SU(2) non-linear $\sigma$
 lagrangian involving pions and nucleons, we will concentrate in the
 pion-nucleon invariant mass $W \le 1.4$ GeV region. For larger
 invariant masses, the chiral expansion will not work, or at least the
 lowest order used in Ref.~\cite{HNV07} will not be sufficient and the
 effect of heavier resonances will become much more
 important~\cite{LPP06}. For this reason in the figure, we always plot
 cross sections calculated with an invariant pion-nucleon mass cut of
 $1.4$ GeV, it is to say we compute $\int_{m_\pi+M}^{1.4~{\rm GeV}}
 dW\, d\sigma/dW $. Such cut is commonly used in the CC pion
 production experimental analyses at intermediate energies. See for
 instance~\cite{anl} where it can also be seen that up to incoming neutrino
 LAB energies of the order of 1 GeV the implementation of this cut
 hardly changes the measured cross section. In the figure, we also
 display low energy data measured in the Argonne National Laboratory
 (ANL)~\cite{anl1NC} for the $\nu n \to \nu p \pi^-$ reaction, which
 do not include any cut in $W$. As shown in Ref.~\cite{HNV07},
 including or not in the theoretical calculation the cut in $W$  only
 influences significantly the result for the highest energy
 data--point, leading to variations of the order of 20\%, which are around 
 half  the experimental error.

Results of Fig.~\ref{fig:sectot} are really significant since the
differences between neutrino and antineutrino induced reactions are always
large, about a factor of two, in all physical channels. This confirms
that this observable can be used to distinguish neutrino from
antineutrino above the pion production threshold.

\begin{figure}[tbh]
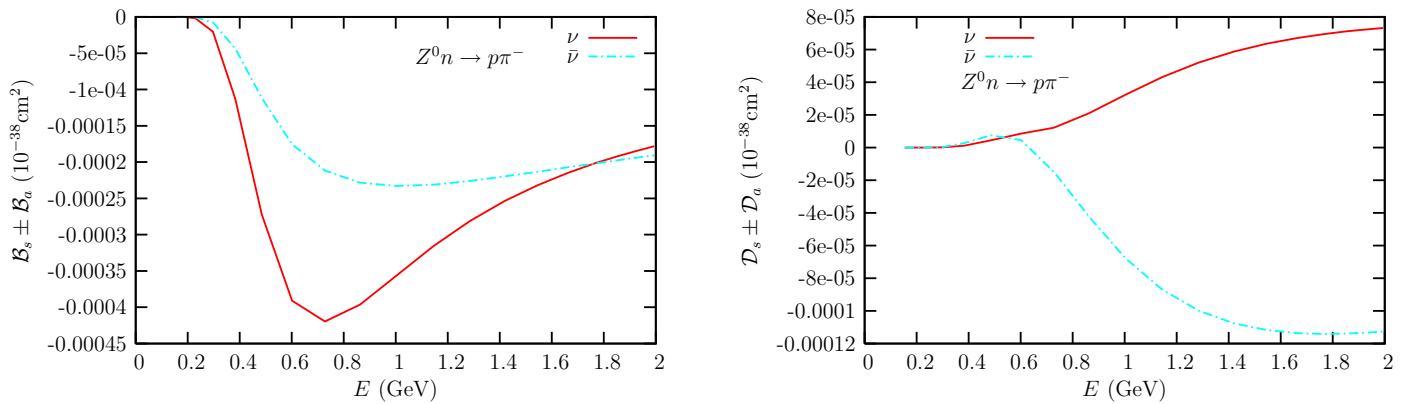

\begin{center}
\makebox[0pt]{\includegraphics[scale=0.7]{B-ppim.epsi}\hspace{1cm}\includegraphics[scale=0.7]{D-ppim.epsi}}
\end{center}
\caption{\footnotesize Predictions of the chiral model of
  Ref.~\cite{HNV07} for the ${\cal B}_s \pm {\cal B}_a$ and ${\cal
  D}_s \pm {\cal D}_a$ (Eqs.~(\ref{eq:int-phisy}) and
  (\ref{eq:int-phiasy})) response functions, as function of the
  neutrino or antineutrino incoming energy.}\label{fig:BD}
\end{figure}

Next we show results, within this model, for ${\cal B}_s \pm {\cal
  B}_a$ and ${\cal D}_s \pm {\cal D}_a$ defined in
  Eqs.~(\ref{eq:int-phisy}) and (\ref{eq:int-phiasy}). In
  Fig.~\ref{fig:BD} we plot these response functions for a $p\pi^-$
  final state. The rest of channels lead to similar results.  Clear
  neutrino-antineutrino asymmetries can be appreciated in both panels
  of the figure. In particular ${\cal D}_s + {\cal D}_a$ and ${\cal
  D}_s - {\cal D}_a$ have opposite signs. The difference between the
  number of events for which the pion comes out above the neutrino
  plane ($N_{\rm abv}$) and those where the pion comes out below this
  plane ($N_{\rm blw}$) is 4 $({\cal D}_s + {\cal D}_a)$ [4 $({\cal
  D}_s -{\cal D}_a)$] for a neutrino [antineutrino] induced
  process. Thus, the sign of this difference would determine whether
  one has a neutrino or an antineutrino beam. Nevertheless, we should
  mention that the neutrino-antineutrino asymmetry based in the total
  cross section discussed above (Fig.~\ref{fig:sectot}) would provide a
  signal with much better statistical significance. This is because
  the ratio $\frac{N_{\rm abv}-N_{\rm blw} }{N_{\rm abv}+N_{\rm blw}}$
  would be of the order of $10^{-2}$ or smaller. Asymmetries based on
  the ${\cal B}$ response function will not be as much unfavorable,
  from the statistical point of view.

 These results point out that the neutrino--antineutrino asymmetries
 based on the total cross sections are certainly much more useful than
 those based on the the pion azimuthal response functions, for
 intermediate neutrino energies. The situation could be different for energies
 larger than those explored with the model of Ref.~\cite{HNV07}.

\begin{acknowledgments}
We thank M.J. Vicente-Vacas for useful discussions. JN acknowledges
the hospitality of the School of Physics \& Astronomy at the
University of Southampton. This work was supported by DGI and FEDER
funds, under contracts FIS2005-00810, BFM2003-00856 and FPA2004-05616,
by Junta de Andaluc\ii a and Junta de Castilla y Le\'on under
contracts FQM0225 and SA104/04, and it is a part of the EU Integrated
Infrastructure Initiative Hadron Physics Project contract
RII3-CT-2004-506078.

\end{acknowledgments}

\appendix
\section{Partially integrated azimuthal distribution}
\label{app:piad}
To define an azimuthal pion distribution, by integrating the outgoing
neutrino variables and pion polar angle in
Eqs.~(\ref{eq:nuphi}) and~(\ref{eq:nubarphi}), requires for each event
a rotation to guarantee that $\vec{q}$ is  in the $Z-$direction
and that $\vec{k}$ and $\vec{k}^{\prime\,}$ are contained in
$XZ-$plane.  The totally integrated unpolarized neutrino cross section
reads
\begin{equation}
\sigma_{\nu}=
\frac{G^2}{16\pi^2 |\vec{k}~|} 
 \int \frac{d^3k^\prime}{|\vec{k}^{\,\prime}|} \frac{d^3k_\pi}{E_\pi} 
 L_{\mu\sigma}^{(\nu)}(k,k^\prime) W^{\mu\sigma}(p,q,k_\pi) 
\end{equation}
To perform the above integrals, we take a fix system of coordinates
${\cal XYZ}$, and since
$L_{\mu\sigma}^{(\nu)}(k,k^\prime)W^{\mu\sigma}(p,q,k_\pi)$ is a
 scalar under spatial rotations, it can be evaluated as 
\begin{equation}
L_{\mu\sigma}^{(\nu)}(Rk,Rk^\prime)W^{\mu\sigma}(Rp,Rq,Rk_\pi)
\end{equation}
where for each $\vec{k}^{\,\prime}$, $R$ is the spatial rotation which
brings $\vec{q}$ and $\vec{k} \times \vec{k}^\prime$ into the ${\cal Z}-$ and
${\cal Y}-$axes, respectively. Such rotation depends on $\vec{k}^\prime$ but
it is independent of $\vec{k}_\pi$. Then
\begin{eqnarray}
\sigma_{\nu}&=& \frac{G^2}{16\pi^2 |\vec{k}~|} \int
    \frac{d^3k^\prime}{|\vec{k}^{\,\prime}|} \frac{d^3k_\pi}{E_\pi}
    L_{\mu\sigma}^{(\nu)}(Rk,Rk^\prime) W^{\mu\sigma}(Rp,Rq,Rk_\pi)\nonumber\\
    &=&\frac{G^2}{16\pi^2 |\vec{k}~|} \int
    \frac{d^3k^\prime}{|\vec{k}^{\,\prime}|} \frac{d^3k_\pi}{E_\pi}
    L_{\mu\sigma}^{(\nu)}(Rk,Rk^\prime) W^{\mu\sigma}(Rp,Rq,k_\pi)
\end{eqnarray}
where in the last step, for a fix $\vec{k}^{\,\prime}$, we have made a
change of variables in the pion integral, $\vec{k}_\pi \to
\vec{k}^R_\pi \equiv R\vec{k}_\pi$, and for simplicity we have called
the dummy variable $\vec{k}^R_\pi$ again $\vec{k}_\pi$. Finally using spherical
coordinates $\vec{k}_\pi = |\vec{k}_\pi|(\sin\theta_\pi
\cos\phi_\pi, \sin\theta \sin\phi_\pi, \cos\theta_\pi )$ for the pion
integral, we get
\begin{eqnarray}
\frac{d\sigma_{\nu}}{d\phi_\pi}&=&
\frac{G^2}{16\pi^2 |\vec{k}~|} 
 \int \frac{d^3k^\prime}{|\vec{k}^{\,\prime}|} \frac{d|\vec{k}_\pi||\vec{k}_\pi|^2 d(\cos\theta_\pi)}{E_\pi} 
 L_{\mu\sigma}^{(\nu)}(Rk,Rk^\prime) W^{\mu\sigma}(Rp,Rq,k_\pi) \nonumber\\
&=&
\frac{G^2}{16\pi^2 |\vec{k}~|} 
 \int \frac{d^3k^\prime}{|\vec{k}^{\,\prime}|} d(\cos\theta_\pi) 
\Big \{ A_s+A_a +
    (B_s+B_a) \cos\phi_\pi + C_s \cos 2\phi_\pi  +(D_s+D_a) \sin\phi_\pi
+ E_s \sin 2\phi_\pi \Big \}\nonumber\\
&=& {\cal A}_s+{\cal A}_a +
    ({\cal B}_s+{\cal B}_a) 
\cos\phi_\pi + {\cal C}_s \cos 2\phi_\pi +({\cal D}_s+{\cal D}_a) 
\sin\phi_\pi +  {\cal E}_s \sin 2\phi_\pi
\end{eqnarray}
since we have $R\vec{q}$ in the ${\cal Z}-$direction and $R\vec{k}$
and $R \vec{k}^{\,\prime}$ are contained in ${\cal XZ}-$plane, and
thus the results of Eq.~(\ref{eq:nuphi}) can be used. In the above
equations, $\phi_\pi$ is the azimuthal angle of the rotated pion
momentum in the ${\cal XYZ}$ system of coordinates, or equivalently, it
is the angle formed by the neutrino and pion planes. The first of
these planes is determined by the incoming ($\vec{k}$\,) and outgoing
($\vec{k}^{\,\prime}$) neutrino momenta, while the transferred
($\vec{q}$\,) and pion ($\vec{k}_\pi$) momenta fix the latter
one. Thus, $\phi_\pi$ is the angle formed by the vectors $\Big(
(\vec{k}\times \vec{k}^{\,\prime})\times \vec{q}\Big )$ and
$\Big(\vec{k}_\pi
-(\vec{k}_\pi\cdot\vec{q})\,\vec{q}/|\vec{q}\,|^2\Big)$, and it is
therefore invariant under spatial rotations.

The antineutrino cross section can be obtained in a similar way, and
thus it follows the neutrino--antineutrino asymmetry relations given
in Eqs.~(\ref{eq:int-phisy}) and~(\ref{eq:int-phiasy}) similar to
those in Eqs.~(\ref{eq:phisy}) and~(\ref{eq:phiasy}), but for
integrated cross sections.


\begin{thebibliography}{blabla}


\bibitem{HHW02} T. Hattori, T. Hasuike, S. Wakaizumi, Phys. Rev. {\bf
  D65} (2002) 073027.

\bibitem{BRJK03} R. Buras, M. Rampp, H.T. Janka, and K. Kifondis,
  Phys. Rev. Lett. {\bf 90} (2003) 241101.

\bibitem{JVRH04} N. Jachowicz, K. Vantournhout, J. Ryckebusch, and
  K. Heyde, Phys. Rev. Lett. {\bf 93} (2004) 082501.

\bibitem{HNV07} E. Hern\'andez, J. Nieves and M. Valverde, {\it
  hep-ph/0701149}.


\bibitem{KLS68} G. Karpman, R. Leonardi and F. Strocchi,
Phys. Rev. {\bf 174} (1968) 1957.


\bibitem{CLS70} F. Cannata, R. Leonardi and F. Strocchi,
Phys. Rev. {\bf D1} (1970) 191.

\bibitem{Adler} S.L. Adler, Ann. Phys. {\bf 50} (1968) 189.


\bibitem{LPP06} O. Lalakulich, E.A. Paschos and G. Piranishvili,
  Phys. Rev. {\bf D74} (2006) 014009.

\bibitem{anl} G.M. Radecky et al.,  Phys. Rev. {\bf  D25} (1982) 1161.

\bibitem{anl1NC} M. Derrick et al., 
Phys. Lett. {\bf B92} (1980) 363;  Erratum-ibid. {\bf B95} (1980) 461.


\end{thebibliography}
\end{document}